\documentclass[10pt]{article}
\usepackage[OE]{express}

\begin{document}

\title{Entanglement Dynamics for Two Spins in an Optical Cavity ---Field
Interaction Induced Decoherence and Coherence Revival}
\author{Xue-Min Bai,\authormark{1} Chun-Ping Gao,\authormark{1} Jun-Qi Li,%
\authormark{1,2} Ni Liu,\authormark{1,3} J.-Q. Liang,\authormark{1,*}}

\address{\authormark{1}Institute of Theoretical Physics, Shanxi University, Taiyuan, Shanxi 030006, China\\
\authormark{2}ljqsxu@163.com\\
\authormark{3}liuni2011520@sxu.edu.cn} 
\email{\authormark{*}jqliang@sxu.edu.cn}

\begin{abstract}
We in this paper study quantum correlations for two neutral spin-particles
coupled with a single-mode optical cavity through the usual magnetic
interaction. Two-spin entangled states for both antiparallel and parallel
spin-polarizations are generated under the photon coherent-state assumption.
Based on the quantum master equation we derive the time-dependent quantum
correlation of Clauser-Horne-Shimony-Holt (CHSH) type explicitly in
comparison with the well known entanglement-measure concurrence. In the
two-spin singlet state, which is recognized as one eigenstate of the system,
the CHSH correlation and concurrence remain in their maximum values
invariant with time and independent of the average photon-numbers either.
The correlation varies periodically with time in the general
entangled-states for the low average photon-numbers. When the photon number
increases to a certain value the oscillation becomes random and the
correlations are suppressed below the Bell bound indicating the decoherence
of the entangled states. In the high photon-number limit the coherence
revivals periodically such that the CHSH correlation approaches the upper
bound value at particular time points associated with the cavity-field
period.
\end{abstract}


\section{Introduction}

The non-locality is one of the most peculiar characteristics of quantum
mechanics without classical correspondence. As a typical example of
non-locality the two-particle entangled-state has become the essential
ingredients in quantum information and computation \cite{Nielsen,
Branciard1, Groblacher}, although it was originally considered by Einstein,
Podolsky, and Rosen to question the completeness of quantum mechanics. In
general, an entangled state for a composite system is not able to be
factorized into a direct product of the individual states of subsystems.
Bell's inequality (BI) derived in terms of classical statistics is a
criteria of the local correlation for the classical bipartite-model. It
actually provides a quantitative test of the entangled states, which are
fundamentally different from the classical world \cite{Paterek, Pironio,
Rabelo, Jaeger, Branciard2, Eisaman, Paternostro, Lee}, while the underlying
physics is obscure \cite{Pawlowski}. The BI and its violation in agreement
with quantum mechanical predictions \cite{Weihs, Aspect, Tittel, Rowe, Wei,
Kwiat, Sakai, Dada} have attracted considerable attentions both
theoretically and experimentally in recent years \cite{Barrett, Zhang,
Kofman, Li, Altintas1, Mazzola1, Derkacz, Pandya, Semenov, Gumberidze,
Werner}. Indeed, the violation of BI has been verified undoubtedly in
various systems, for example, with entangled photons \cite{Weihs, Aspect,
Tittel}, trapped ions \cite{Rowe} and switchable Josephson qubits \cite{Wei}%
. Soon after the Bell's pioneer work the original BI was modified to various
new forms \cite{Clauser, Bierhorst, Leggett, Hensen2015, Quintino, Dutta,
Roy}, among which a more suitable inequality for the quantitative test was
formulated by Clauser-Horne-Shimony-Holt (CHSH). The entanglement and BI
have been also investigated in optical cavity with identical atoms \cite%
{Zhen, Chen1, Li2005, Xu, Stace, Wang, Gong, Francica}. Recently a loop-free
experiment on the violation of BI was reported \cite{Waldherr, Hensen} using
the spins of a nitrogen-vacancy defect in diamond \cite{Chen, Zhao1, Zhao2,
Yin, Scala, Kolkowitz}.

Nonetheless, most of these current studies are focused on the measurement of
static entangled states. It is of fundamental importance to explore the
dynamic behavior of quantum correlation for the bipartite entangled states
and the related violation of BI in an interacting system. The environment
induced decoherence is considered as one of the most serious obstacle to
realize the quantum operation. The dynamic evolution of quantum correlations
for a two-qubit system has been studied extensively in various
dissipative-environments\ \cite{Altintas2, Sun, Lu, Werlang, Hao, Yuan}.
Motivated by the experimental test of BI in spin systems \cite{Waldherr,
Hensen, Polozova, Ansman} we in the present paper study a dynamic model for
two spins interacting with a cavity-field to see the field-interaction
effect on the entanglement. We demonstrate that general entangled states
indeed can be generated and then derive the quantum correlation in terms of
quantum probability statistics under the assumption of measuring
outcome-independence \cite{Song}. The quantum CHSH-correlation denoted by $%
P_{CHSH}$ results generally in the violation of BI, $P_{CHSH}<2$, for the
entangled states, and the maximum correlation-value is found as $%
P_{CHSH}^{\max }=2\sqrt{2}$. The dynamic evolution of the maximum quantum
CHSH-correlation $P_{CHSH}^{\max }$ is evaluated explicitly with the help of
quantum master equation and the result is compared with the entanglement
measure, concurrence. We reveal the field-interaction induced decoherence
and the coherence revival, which is characterized by the violation of BI.

The paper is organized as follows: we present the model Hamiltonian and
stationary solutions in Sec.II. The general entangled states are generated
in the spin cavity system. In Sec. III, we derive the quantum
CHSH-correlation in terms of quantum statistics for the two-spin entangled
states. The quantum dynamics of density operator is formulated in Sec.IV.
The time-evolutions of $P_{CHSH}^{\max }$ and concurrence are presented in
Sec.V. Finally, we in Sec.VI summarize and discuss the results.

\section{Stationary solutions and entangled states}

Following the recent experiment for the violation of BI with
nitrogen-vacancy defects in diamond \cite{Waldherr, Hensen, Chen, Zhao1} we
assume that the two spins are subject to a one-mode optical cavity of
frequency $\omega $. The usual Zeeman energy for two spins in the magnetic
component of quantized light field gives rise to the Hamiltonian 
\begin{equation}
H=\omega a^{\dagger }a+ig\sum_{i=1}^{2}(a\sigma _{i}^{+}-a^{\dagger }\sigma
_{i}^{-}),  \label{Eq1}
\end{equation}%
with the convention $\hbar =1$ throughout the paper. The Pauli matrices $%
\sigma _{i}^{z}$, $\sigma _{i}^{\pm }=\sigma _{i}^{x}\pm i\sigma _{i}^{y}$
are defined to describe the spin operators. $a$ ($a^{\dagger }$) is the
annihilation (creation) operator of the single-mode cavity field.

We are looking for the stationary solutions of Hamiltonian Eq. (\ref{Eq1})
in the optical coherent state $\left\vert \alpha \right\rangle $ under the
semiclassical approximation. Taking the average in the coherent state $%
\left\vert \alpha \right\rangle $, in which the complex eigenvalue of the
photon annihilation operator, that $a|\alpha \rangle =\alpha |\alpha \rangle 
$, is parameterized as $\alpha =\gamma e^{i\phi }$, we have an effective
Hamiltonian with spin operator only 
\begin{equation}
H_{sp}(\alpha )=\left\langle \alpha \right\vert H\left\vert \alpha
\right\rangle =\omega \gamma ^{2}+\sum_{i=1}^{2}i\gamma g[e^{i\phi }\sigma
_{i}^{+}-e^{-i\phi }\sigma _{i}^{-}].  \label{Eq2}
\end{equation}%
$\gamma ^{2}=\langle \alpha |a^{\dagger }a|\alpha \rangle $ denotes the
average photon-number, which plays a important role in the entanglement
dynamics. The spin Hamiltonian $H_{sp}(\alpha )$ can be diagonalized in the
two-qubit bases $|e_{1}\rangle =\left\vert +,+\right\rangle $, $%
|e_{2}\rangle =\left\vert +,-\right\rangle $, $|e_{3}\rangle =\left\vert
-,+\right\rangle $, $|e_{4}\rangle =\left\vert -,-\right\rangle $ with the
energy eigenvalues given by%
\begin{eqnarray}
\varepsilon _{0}(\gamma ) &=&\omega \gamma ^{2}-2g\gamma ,  \nonumber \\
\varepsilon _{3}(\gamma ) &=&\omega \gamma ^{2}+2g\gamma ,  \nonumber \\
\varepsilon _{1}(\gamma ) &=&\varepsilon _{2}(\gamma )=\omega \gamma ^{2},
\label{Eq3}
\end{eqnarray}%
which are functions of parameter $\gamma $. The corresponding eigenstates
are seen to be 
\begin{eqnarray}
\left\vert \psi _{0}\right\rangle &=&\frac{1}{2}[e^{-i\phi }\left\vert
-,-\right\rangle -e^{i\phi }\left\vert +,+\right\rangle -i(\left\vert
+,-\right\rangle +\left\vert -,+\right\rangle )],  \nonumber \\
\left\vert \psi _{3}\right\rangle &=&\frac{1}{2}[e^{-i\phi }\left\vert
-,-\right\rangle -e^{i\phi }\left\vert +,+\right\rangle +i(\left\vert
+,-\right\rangle +\left\vert -,+\right\rangle )],  \nonumber \\
\left\vert \psi _{2}\right\rangle &=&\frac{1}{\sqrt{2}}(e^{i\phi }\left\vert
+,+\right\rangle +e^{-i\phi }\left\vert -,-\right\rangle ),  \nonumber \\
\left\vert \psi _{1}\right\rangle &=&\frac{1}{\sqrt{2}}(\left\vert
-,+\right\rangle -\left\vert +,-\right\rangle ),  \label{Eq4}
\end{eqnarray}

The two-spin entangled states indeed can be generated in the optical cavity
under the semiclassical field approximation. $\left\vert \psi
_{1}\right\rangle $ is the well studied two-spin singlet state in relation
with the nonlocality and BI. While $\left\vert \psi _{2}\right\rangle $,
which is a degenerate state of $\left\vert \psi _{1}\right\rangle $, is the
entangled state with parallel spin polarization. The stability of the states 
$\left\vert \psi _{0}\right\rangle $, $\left\vert \psi _{1}\right\rangle $,
and $\left\vert \psi _{2}\right\rangle $ can be examined with variational
method regarding $\gamma $ as a variational parameter of energy function.
For example, the solution $\gamma =0$ of the energy extremum equation $%
\partial $ $\varepsilon _{1}(\gamma )/\partial \gamma =0$ is obviously the
energy minimum of the energy function $\varepsilon _{1}(\gamma )$. Thus $%
\left\vert \psi _{1}\right\rangle $, and $\left\vert \psi _{2}\right\rangle $
are stable states of normal phase with zero average photon-number. While $%
\left\vert \psi _{0}\right\rangle $ is the ground state of superradiant
phase \cite{Zhao4, Wang4, Wang5}. We solve in the present paper the dynamic
equation of motion for the given initial entangled states and then study the
time-evolution of measuring outcome-correlations. Particular attention is
paid on the field interaction induced decoherence and coherence revival. For
the sake of simplicity we consider the dynamics associated with the initial
states $\left\vert \psi _{1}\right\rangle $, and $\left\vert \psi
_{2}\right\rangle $.

\section{Measuring outcome correlation and Bell's inequality}

BI derived from classical statistics with the assumption of locality, was
one of the first criteria to detect the quantum entanglement. Recently the
BIs and their violations were revisited in terms of quantum probability
statistics with the help of state density-operator \cite{Song}, which has
advantage to formulate the various forms of inequality and the violation in
a unified manner. We consider a general two-spin entangled state with
antiparallel spin-polarization 
\begin{equation}
\left\vert \psi \right\rangle =\sin \xi e^{i\eta }\left\vert
+,-\right\rangle +\cos \xi e^{-i\eta }\left\vert -,+\right\rangle ,
\label{p01}
\end{equation}%
where $\xi $, $\eta $ are two arbitrary angle-parameters. The density
operator can be separated into the local and non-local parts%
\begin{equation}
\rho _{\psi }=\rho _{\psi }^{lc}+\rho _{\psi }^{nlc}.  \label{sep}
\end{equation}%
The local part 
\begin{equation}
\rho _{\psi }^{lc}=\sin ^{2}\xi \left\vert +,-\right\rangle \langle
+,-|+\cos ^{2}\xi \left\vert -,+\right\rangle \langle -,+|,  \label{lcp}
\end{equation}%
which is the classical probability state, gives rise to BIs. While the
non-local part 
\begin{equation}
\rho _{\psi }^{nlc}=\sin \xi \cos \xi e^{-i2\eta }\left\vert
+,-\right\rangle \langle -,+|+\sin \xi \cos \xi e^{i2\eta }\left\vert
-,+\right\rangle \langle +,-|,  \label{nlcp}
\end{equation}%
which comes from the quantum interference between two components of the
entangled state leads to the violation of BIs. Following Bell the
measurements of two spins are performed independently along arbitrary
directions, say $\mathbf{a}$ and $\mathbf{b}$, respectively. The measuring
outcomes fall into the eigenvalues of projection spin-operators%
\[
\sigma \cdot \mathbf{a}\left\vert \pm \mathbf{a}\right\rangle =\pm
\left\vert \pm \mathbf{a}\right\rangle 
\]%
and 
\[
\sigma \cdot \mathbf{b}\left\vert \pm \mathbf{b}\right\rangle =\pm
\left\vert \pm \mathbf{b}\right\rangle , 
\]%
according to the quantum measurement theory. By solving the above
eigenequations we obtain the explicit form of eigenstates 
\[
\left\vert +\mathbf{n}\right\rangle =\cos (\frac{\theta _{n}}{2})\left\vert
+\right\rangle +\sin (\frac{\theta _{n}}{2})e^{i\Phi _{n}}\left\vert
-\right\rangle 
\]%
and 
\[
\left\vert -\mathbf{n}\right\rangle =\sin (\frac{\theta _{n}}{2})\left\vert
+\right\rangle -\cos (\frac{\theta _{n}}{2})e^{i\Phi _{n}}\left\vert
-\right\rangle , 
\]%
which are known as the spin coherent states \cite{Zhao4, Wang4, Wang5} of
north- and south- pole gauges. Here, $\mathbf{n}=(\sin \theta _{n}\cos \Phi
_{n}$, $\sin \theta _{n}\sin \Phi _{n}$, $\cos \theta _{n})$ with $\mathbf{n}%
=\mathbf{a}$, $\mathbf{b}$ is a unit vector parameterized by the polar and
azimuthal angles ($\theta _{n}$ , $\Phi _{n}$). The outcome-independent base
vectors of two-spin measurements are the product eigenstates of operators $%
\sigma \cdot \mathbf{a}$ and $\sigma \cdot \mathbf{b}$ labeled arbitrarily
as 
\begin{equation}
\left\vert 1\right\rangle \!=\!\left\vert \!+\mathbf{a,\!+b}\right\rangle
,\!\left\vert 2\right\rangle \!=\!\left\vert \!+\mathbf{a,\!-b}\right\rangle
,\!\left\vert 3\right\rangle \!=\!\left\vert \!-\mathbf{a,\!+b}\right\rangle
,\!\left\vert 4\right\rangle \!=\!\left\vert \!-\mathbf{a,\!-b}\right\rangle
.  \label{ab}
\end{equation}%
The measuring outcome correlation-probability for two spins respectively
along directions $\mathbf{a}$ and $\mathbf{b}$ is thus obtained in terms of
the quantum statistical-average under the outcome-independent base vectors
Eq. (\ref{ab}) 
\begin{equation}
P(a,b)=Tr[(\sigma \cdot \mathbf{a})(\sigma \cdot \mathbf{b})\rho _{\psi }].
\label{pab}
\end{equation}%
Using the separated density operators Eq. (\ref{sep}) the measuring outcome
correlation-probability Eq. (\ref{pab}) can be also split into the local and
non-local parts 
\[
P(a,b)=P_{lc}(a,b)+P_{nlc}(a,b). 
\]%
The local part Eq. (\ref{lcp}) gives rise to the Bell correlation \cite{Song}
of two-direction ($a,b$) measurements 
\begin{equation}
P_{lc}(a,b)=-\cos \theta _{a}\cos \theta _{b}.  \label{co}
\end{equation}%
Substituting the the Bell correlation Eq. (\ref{co}) into the
CHSH-correlation-probability of four-direction measurements defined as 
\begin{equation}
P_{CHSH}=\left\vert P(a,b)+P(a,c)+P(d,b)-P(d,c)\right\vert ,  \label{CHSH}
\end{equation}%
we recover the well known CHSH-inequality 
\begin{equation}
P_{CHSH}^{lc}\!=\!\left\vert \cos \theta _{a}(\cos \theta _{b}\!+\!\cos
\theta _{c})\!+\!\cos \theta _{d}(\cos \theta _{b}\!-\!\cos \theta
_{c})\right\vert \!\leq 2,  \label{lc}
\end{equation}%
which was derived originally by CHSH from classical statistics with the
assumption of locality. The CHSH-inequality Eq. (\ref{lc}) is valid for
arbitrary entangled states given by Eq. (\ref{p01}) considering the local
part Eq. (\ref{lcp}) only. Including the non-local part Eq. (\ref{nlcp}) the
quantum correlation for the two-direction measurements is simply a scaler
product of the two unit-vectors $\mathbf{a}$ and $\mathbf{b}$ 
\begin{equation}
P(a,b)=-\mathbf{a\cdot b,}  \label{qtu}
\end{equation}%
which is valid, however, under the condition $\xi =3\pi /4$, $\eta =n\pi $
with $n$ being a integer. Thus the arbitrary entangled state Eq. (\ref{p01})
reduces to the two-spin singlet state $\left\vert \psi _{1}\right\rangle $
in Eq. (\ref{Eq4}). With the quantum correlation Eq. (\ref{qtu}) the quantum
CHSH-probability Eq. (\ref{CHSH}) becomes \cite{Song,Zhang4} 
\begin{equation}
P_{CHSH}=\left\vert \mathbf{a\cdot (b+c)+d\cdot (b-c)}\right\vert .
\label{Eq15}
\end{equation}%
When the unit vector $\mathbf{b}$ is perpendicular to $\mathbf{c}$, and $%
\mathbf{a}$, $\mathbf{d}$ are respectively parallel to ($\mathbf{b+c}$), ($%
\mathbf{b-c}$), we obtain the maximum quantum CHSH-probability as 
\begin{equation}
P_{CHSH}^{\max }=2\sqrt{2},  \label{max}
\end{equation}%
which indicates the maximum violation of CHSH-inequality Eq. (\ref{lc}). Our
formalism shows a direct relation between the decoherence and
CHSH-inequality, which can serves a criteria of coherence for the two parts
of entangled state. Most recently it was proved that for the entangled state
of parallel spin polarization%
\begin{equation}
\left\vert \psi \right\rangle =\sin \xi e^{i\eta }\left\vert
+,+\right\rangle +\cos \xi e^{-i\eta }\left\vert -,-\right\rangle ,
\label{bbbb}
\end{equation}%
the Bell correlation of local part is simply \cite{Zhang4}%
\[
P_{lc}(a,b)=\cos \theta _{a}\cos \theta _{b}, 
\]%
which is different from that of antiparallel spin-polarization Eq. (\ref{co}%
) only by a sign change. The CHSH-inequality for the local part Eq. (\ref{lc}%
) remains not changed. The quantum CHSH-correlation-probability Eq. (\ref%
{max}) is valid for the entangled state under the condition $\xi =\pi /4$, $%
\eta =n\pi $, namely, $\left\vert \psi \right\rangle =1/2(\left\vert
+,+\right\rangle +\left\vert -,-\right\rangle ).$

We are going to study the dynamics of CHSH-probability to reveal the field
interaction induced effect on the entanglement. The dynamic evolution of $%
P_{CHSH}$ can be obtained from the time dependent density-operator $\rho (t)$
in terms of the quantum probability statistics presented in this paper. For
speciality we consider only the maximum quantum CHSH-probability Eq. (\ref%
{max}), which has been shown to evaluate in a clever way \cite{Horodecki1,
Horodecki2} directly from the state-density operator itself.

\section{Quantum dynamics with photon coherent state}

The dynamic evolution of two-qubit entangled states in the optical cavity
can be evaluated from the Schr$\ddot{o}$dinger equation $i\partial /\partial
t|\psi \rangle =H|\psi \rangle $. To this end we begin with the interaction
picture $i\partial /\partial t|\psi _{i}\rangle =H_{i}|\psi _{i}\rangle $,
where $|\psi _{i}\rangle =R(t)|\psi \rangle $ and the unitary operator is 
\[
R(t)=e^{i\omega a^{\dagger }at}. 
\]%
The interaction-picture Hamiltonian is seen to be 
\begin{eqnarray}
H_{i} &=&R(t)HR^{\dag }(t)-iR(t)\frac{\partial }{\partial t}R^{\dag }(t) 
\nonumber \\
&=&ig\sum_{i=1}^{2}(ae^{-i\omega t}\sigma _{i}^{+}-a^{\dagger }e^{i\omega
t}\sigma _{i}^{-}).  \label{Eq18}
\end{eqnarray}%
The entail time-evolution operator of the system reads as 
\begin{equation}
U(t)=R^{\dag }(t)U_{i}(t),  \label{unity}
\end{equation}%
with $U_{i}(t)=\exp (-iH_{i}t)$ being the time evolution operator for the
interaction-picture Hamiltonian Eq. (\ref{Eq1}). In the two-qubit bases $%
|e_{i}\rangle $ ($i=1,2,3,4$), time-evolution operator possesses a matrix
form \cite{Kim}, 
\begin{equation}
U_{i}(t)\!=\left( 
\begin{array}{cccc}
1\!-\!2a\frac{\sin ^{2}(gt\sqrt{\frac{S}{2}})}{S}a^{\dagger } & a\frac{\sin
(gt\sqrt{2S})}{\sqrt{2S}} & a\frac{\sin (gt\sqrt{2S})}{\sqrt{2S}} & 2a\frac{%
\sin ^{2}(gt\sqrt{\frac{S}{2}})}{S}a \\ 
-\frac{\sin (gt\sqrt{2S})}{\sqrt{2S}}a^{\dagger } & \cos ^{2}(gt\sqrt{\frac{S%
}{2}}) & -\sin ^{2}(gt\sqrt{\frac{S}{2}}) & \frac{\sin (gt\sqrt{2S})}{\sqrt{%
2S}}a \\ 
-\frac{\sin (gt\sqrt{2S})}{\sqrt{2S}}a^{\dagger } & -\sin ^{2}(gt\sqrt{\frac{%
S}{2}}) & \cos ^{2}(gt\sqrt{\frac{S}{2}}) & \frac{\sin (gt\sqrt{2S})}{\sqrt{%
2S}}a \\ 
2a^{\dagger }\frac{\sin ^{2}(gt\sqrt{\frac{S}{2}})}{S}a^{\dagger } & 
-a^{\dagger }\frac{\sin (gt\sqrt{2S})}{\sqrt{2S}} & -a^{\dagger }\frac{\sin
(gt\sqrt{2S})}{\sqrt{2S}} & 1-2a^{\dagger }\frac{\sin ^{2}(gt\sqrt{\frac{S}{2%
}})}{S}a%
\end{array}
\right),
\end{equation}
where $S=2a^{\dagger}a+1$.

From the quantum master equation 
\begin{equation}
i\frac{d\rho (t)}{dt}=[H,\rho (t)],  \label{Eq20}
\end{equation}%
the density operator at time $t$ is found as $\rho (t)=U(t)\rho
(0)U^{\dagger }(t)$, where the initial density operator 
\[
\rho (0)=\rho _{\psi }(0)\rho _{f}(0), 
\]%
is the product of the spin part $\rho _{\psi }(0)=\left\vert \psi
(0)\right\rangle \langle \psi (0)|$ and the cavity-field part $\rho
_{f}(0)=|\alpha \rangle \langle \alpha |$ with $|\alpha \rangle $ being the
photon coherent state \cite{Zhao4}. We are interested in the
field-interaction effect on the quantum measuring-outcome correlation and
thus assume the initial state being normalized two-spin entangled-state Eq. (%
\ref{p01}). The matrix representation of the density operator Eq. (\ref{p01}%
) in the two-qubit bases is 
\begin{equation}
\rho _{\psi }(0)=\left( 
\begin{array}{cccc}
0 & 0 & 0 & 0 \\ 
0 & \sin ^{2}\xi & \sin \xi \cos \xi e^{i2\eta } & 0 \\ 
0 & \sin \xi \cos \xi e^{-i2\eta } & \cos ^{2}\xi & 0 \\ 
0 & 0 & 0 & 0%
\end{array}%
\right) .  \label{Eq21}
\end{equation}%
Using the explicit form of photon coherent state in the Fock space, $|\alpha
\rangle =\sum_{n}\alpha ^{n}/\sqrt{n!}\exp \{-|\alpha |^{2}/2\}|n\rangle $ ,
we take the trace of density operator over the photon number state $%
|n\rangle $ to obtain the reduced density operator that 
\[
\rho _{r}(t)=\sum\limits_{n=0}^{\infty }\rho _{n}(t),\qquad \rho
_{n}(t)=\langle n|\rho (t)|n\rangle , 
\]%
with%
\begin{equation}
\rho _{n}(t)=\sum\limits_{j,l=0}^{\infty }\langle n|U(t)|j\rangle \rho
_{\psi }(0)\langle l|U^{\dagger }(t)|n\rangle \frac{\gamma
^{j+l}e^{i(j-l)\phi }}{\sqrt{j!l!}}e^{-\gamma ^{2}}.  \label{Eq22}
\end{equation}%
The reduced density matrix is written in the two-qubit bases explicitly as 
\begin{equation}
\rho _{r}(t)=\sum\limits_{n=0}^{\infty }\left( 
\begin{array}{cccc}
(\rho _{n})_{11} & (\rho _{n})_{12} & (\rho _{n})_{13} & (\rho _{n})_{14} \\ 
(\rho _{n}^{\ast })_{12} & (\rho _{n})_{22} & (\rho _{n})_{23} & (\rho
_{n})_{24} \\ 
(\rho _{n}^{\ast })_{13} & (\rho _{n}^{\ast })_{23} & (\rho _{n})_{33} & 
(\rho _{n})_{34} \\ 
(\rho _{n}^{\ast })_{14} & (\rho _{n}^{\ast })_{24} & (\rho _{n}^{\ast
})_{34} & (\rho _{n})_{44}%
\end{array}%
\right) .  \label{den}
\end{equation}%
The diagonal elements of the density matrix $\rho _{n}(t)$ are seen to be 
\begin{eqnarray*}
(\rho _{n})_{11} &=&\langle e_{1}|\rho _{n}|e_{1}\rangle =\frac{e^{-\gamma
^{2}}}{n!}C_{1}^{2}\left( \sin 2\xi \cos 2\eta +1\right) , \\
(\rho _{n})_{22} &=&\frac{e^{-\gamma ^{2}}}{n!}(C_{2}^{2}\sin ^{2}\xi
+C_{3}^{2}\cos ^{2}\xi +C_{2}C_{3}\sin 2\xi \cos 2\eta ), \\
(\rho _{n})_{33} &=&\frac{e^{-\gamma ^{2}}}{n!}(C_{2}^{2}\cos ^{2}\xi
+C_{3}^{2}\sin ^{2}\xi +C_{2}C_{3}\sin 2\xi \cos 2\eta ), \\
(\rho _{n})_{44} &=&\frac{e^{-\gamma ^{2}}}{n!}C_{4}^{2}\left( \sin 2\xi
\cos 2\eta +1\right) ,
\end{eqnarray*}%
and the off-diagonal elements are 
\begin{eqnarray*}
(\rho _{n})_{12} &=&\frac{e^{-\gamma ^{2}}}{n!}C_{1}e^{-2i\eta }(C_{2}\sin
\xi +C_{3}e^{2i\eta }\cos \xi )(e^{2i\eta }\sin \xi +\cos \xi ), \\
(\rho _{n})_{13} &=&\frac{e^{-\gamma ^{2}}}{n!}C_{1}e^{-2i\eta
}(C_{2}e^{2i\eta }\cos \xi +C_{3}\sin \xi )(e^{2i\eta }\sin \xi +\cos \xi ),
\\
(\rho _{n})_{14} &=&\frac{e^{-\gamma ^{2}}}{n!}C_{1}C_{4}\left( \sin 2\xi
\cos 2\eta +1\right) , \\
(\rho _{n})_{23} &=&\frac{e^{-\gamma ^{2}}}{2n!}\left( C_{2}^{2}e^{2i\eta
}\sin 2\xi +2C_{2}C_{3}+C_{3}^{2}e^{-2i\eta }\sin 2\xi \right) , \\
(\rho _{n})_{24} &=&\frac{e^{-\gamma ^{2}}}{n!}C_{4}e^{-2i\eta
}(C_{2}e^{2i\eta }\sin \xi +C_{3}\cos \xi )(e^{2i\eta }\cos \xi +\sin \xi ),
\\
(\rho _{n})_{34} &=&\frac{e^{-\gamma ^{2}}}{n!}C_{4}e^{-2i\eta }(C_{2}\cos
\xi +C_{3}e^{2i\eta }\sin \xi )(e^{2i\eta }\cos \xi +\sin \xi ).
\end{eqnarray*}%
We have ($\rho _{n})_{ij}=(\rho _{n})_{ji}^{\ast }$, since the state-density
operator $\rho _{n}(t)$ is Hermitian. The time functions are given by 
\begin{eqnarray*}
C_{1} &=&\gamma ^{n+1}\frac{\sin (gt\sqrt{4n+6})}{\sqrt{4n+6}}, \\
C_{2} &=&\gamma ^{n}\cos ^{2}(gt\sqrt{\frac{2n+1}{2}}), \\
C_{3} &=&-\gamma ^{n}\sin ^{2}(gt\sqrt{\frac{2n+1}{2}}), \\
C_{4} &=&-\gamma ^{n-1}\frac{n\sin (gt\sqrt{4n-2})}{\sqrt{4n-2}}.
\end{eqnarray*}

For the entangled initial-state of parallel spin polarization Eq. (\ref{bbbb}%
) the density operator becomes 
\begin{equation}
\rho _{\psi }(0)=\left( 
\begin{array}{cccc}
\sin ^{2}\xi & 0 & 0 & \sin \xi \cos \xi e^{2i\eta } \\ 
0 & 0 & 0 & 0 \\ 
0 & 0 & 0 & 0 \\ 
\sin \xi \cos \xi e^{-2i\eta } & 0 & 0 & \cos ^{2}\xi%
\end{array}%
\right) .  \label{Eq24}
\end{equation}%
The diagonal elements of the density matrix $\rho _{n}(t)$ are 
\begin{eqnarray*}
(\rho _{n})_{11} &=&\frac{e^{-\gamma ^{2}}}{n!}(D_{1}^{2}\sin ^{2}\xi
+D_{2}^{2}\cos ^{2}\xi +D_{1}D_{2}\sin 2\xi \cos 2\eta ), \\
(\rho _{n})_{22} &=&\frac{e^{-\gamma ^{2}}}{n!}(D_{3}^{2}\sin ^{2}\xi
+D_{4}^{2}\cos ^{2}\xi +D_{3}D_{4}\sin 2\xi \cos 2\eta ), \\
(\rho _{n})_{44} &=&\frac{e^{-\gamma ^{2}}}{n!}(D_{5}^{2}\sin ^{2}\xi
+D_{6}^{2}\cos ^{2}\xi +D_{5}D_{6}\sin 2\xi \cos 2\eta ), \\
(\rho _{n})_{33} &=&(\rho _{n})_{22},
\end{eqnarray*}%
and the off-diagonal elements read%
\begin{eqnarray*}
(\rho _{n})_{12} &=&\frac{e^{-\gamma ^{2}}}{n!}(D_{4}\cos \xi +e^{-2i\eta
}D_{3}\sin \xi )(D_{2}\cos \xi +e^{2i\eta }D_{1}\sin \xi ), \\
(\rho _{n})_{14} &=&\frac{e^{-\gamma ^{2}}}{n!}(D_{6}\cos \xi +e^{-2i\eta
}D_{5}\sin \xi )(D_{2}\cos \xi +e^{2i\eta }D_{1}\sin \xi ), \\
(\rho _{n})_{24} &=&\frac{e^{-\gamma ^{2}}}{n!}(D_{6}\cos \xi +e^{-2i\eta
}D_{5}\sin \xi )(D_{4}\cos \xi +e^{2i\eta }D_{3}\sin \xi ), \\
(\rho _{n})_{13} &=&(\rho _{n})_{12}, \\
(\rho _{n})_{23} &=&(\rho _{n})_{22}, \\
(\rho _{n})_{34} &=&(\rho _{n})_{24},
\end{eqnarray*}%
with the time functions being 
\begin{eqnarray*}
D_{1} &=&\gamma ^{n}(1-\frac{2(n+1)\sin ^{2}(gt\sqrt{\frac{2n+3\allowbreak }{%
2}})}{2n+3}), \\
D_{2} &=&\gamma ^{n+2}\frac{2\sin ^{2}(gt\sqrt{\frac{2n+3}{2}})}{2n+3}, \\
D_{3} &=&-\gamma ^{n-1}\frac{n\sin (gt\sqrt{4n+2})}{\sqrt{4n+2}}, \\
D_{4} &=&\gamma ^{n+1}\frac{\sin (gt\sqrt{4n+2})}{\sqrt{4n+2}}, \\
D_{5} &=&\gamma ^{n-2}\frac{2n(n-1)\sin ^{2}(gt\sqrt{\frac{2n-1}{2}})}{2n-1},
\\
D_{6} &=&\gamma ^{n}(1-\frac{2n\sin ^{2}(gt\sqrt{\frac{2n-1}{2}})}{2n-1}).
\end{eqnarray*}

In the following we are going to investigate measuring outcome-correlations
of two spins in terms of the density operator $\rho _{r}(t)$ to see the
field-interaction induced effect on the entanglement. The dynamic evolution
of quantum CHSH-correlation-particularly $P_{CHSH}^{\max }$ is evaluated in
comparison with the entanglement measure, concurrence.

\section{Time evolution of quantum correlation}

We now study the entanglement dynamics in terms of the time-evolution of
density matrix $\rho _{r}(t)$, from which the time-dependent maximum quantum
CHSH-correlation-probability is derived. To this end we define \cite%
{Horodecki1, Horodecki2} a matrix relating to $\rho _{r}(t)$ as 
\[
T_{\rho _{r}(t)}=\rho _{r}(t)\sigma \otimes \sigma . 
\]%
A symmetric matrix is given by 
\begin{equation}
U_{\rho _{r}(t)}=T_{\rho _{r}(t)}^{T}T_{\rho _{r}(t)},  \label{UPR}
\end{equation}%
where $T_{\rho }^{T}$\ is the transposition of $T_{\rho }$. The maximum
quantum CHSH-correlation-probability can be found \cite{Horodecki1,
Horodecki2} from the eigenvalues of the symmetric matrix $U_{\rho _{r}(t)}$ 
\begin{equation}
P_{CHSH}^{\max }(t)=2\sqrt{m(\rho _{r})}.  \label{Eq26}
\end{equation}%
where $m(\rho _{_{r}(t)})=\max_{j<k}(u_{j}+u_{k})$\ with $u_{j}$\ $(j=1,2,3)$%
\ being the eigenvalues of $U_{_{\rho _{r}(t)}}$ defined in Eq. (\ref{UPR}).
In the absence of the cavity field we have $P_{CHSH}^{\max }=2\sqrt{m(\rho
_{\psi })}=2\sqrt{2}$, which coincides with Eq. (\ref{max}). The $%
P_{CHSH}^{\max }(t)$-value can be used to indicate the entanglement that
when $P_{CHSH}^{\max }(t)<2$, the entanglement collapses due to the
field-interaction induced quantum decoherence.

It is interesting to compare the $P_{CHSH}^{\max }(t)$ with the concurrence,
which is a well known measure of the entanglement \cite{Wootters} for a
two-qubit system. The instantaneous concurrence is defined by \cite{Wootters}
\begin{equation}
C(t)=\max \{0,\Lambda (t)\},  \label{Eq27}
\end{equation}%
where $\Lambda (t)$ $=$ $\lambda _{1}(t)-\lambda _{2}(t)-\lambda
_{3}(t)-\lambda _{4}(t)$. $\lambda _{i}(t)$ denotes the square root of the
instantaneous eigenvalue of the matrix $\rho _{r}(t)(\sigma _{y}\otimes
\sigma _{y})\rho _{r}^{\ast }(t)(\sigma _{y}\otimes \sigma _{y})$ for $%
i=1,2,3,4$ in the decreasing order of eigenvalue magnitudes. $\rho
_{r}^{\ast }(t)$ is the complex conjugate of the two-qubit density matrix $%
\rho _{r}(t)$.

The maximum CHSH-correlation-probability $P_{CHSH}^{\max }(t)$ is compared
with concurrence $C(t)$, so that to establish a dynamic relation between the
violation of CHSH inequality and the entanglement measure. In the photon
coherent-state the average photon-number $\gamma ^{2}$, which, we will see,
affects greatly the entanglement dynamics.

\subsection{Entangled states of antiparallel spin polarizations}

\begin{figure}[ptb]
\centering\includegraphics[width=13cm]{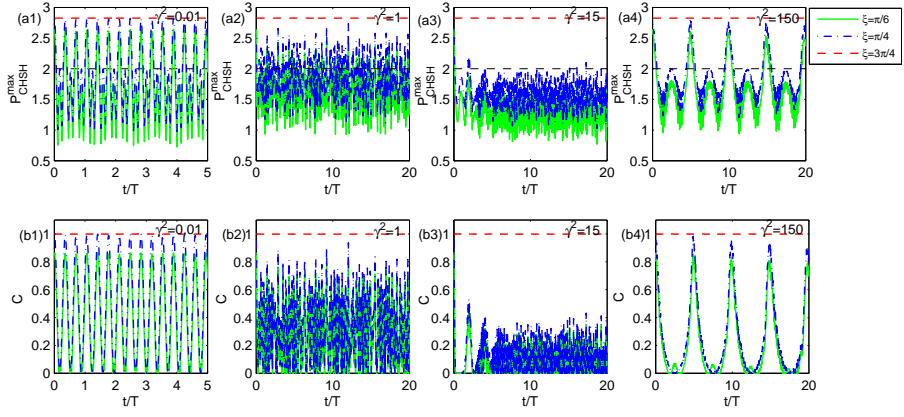}
\caption{{}Color online) The maximum quantum CHSH-probability $%
P_{CHSH}^{\max }(t)$ (a) and concurrence $C(t)$ (b) as functions of time $t$
measured in the period $T$ of optic field for the entangled-state of
antiparallel spin-polarization with angle parameters $\protect\eta =0$ and $%
\protect\xi =\protect\pi /6$ (green solid line), $\protect\pi /4$ (blue dot
and dash line) $\protect\pi /3$ (red dash line) with average photon-number $%
\protect\gamma ^{2}=0.01$ (1), $1$ (2), $15$ (3), $150$ (4). $P_{CHSH}^{\max
}(t)=2\protect\sqrt{2}$, $C(t)=1$ not varying with time for the spin singlet
state $\left\vert \Psi _{1}\right\rangle $.}
\end{figure}

\begin{figure}[ptb]
\centering\includegraphics[width=13cm]{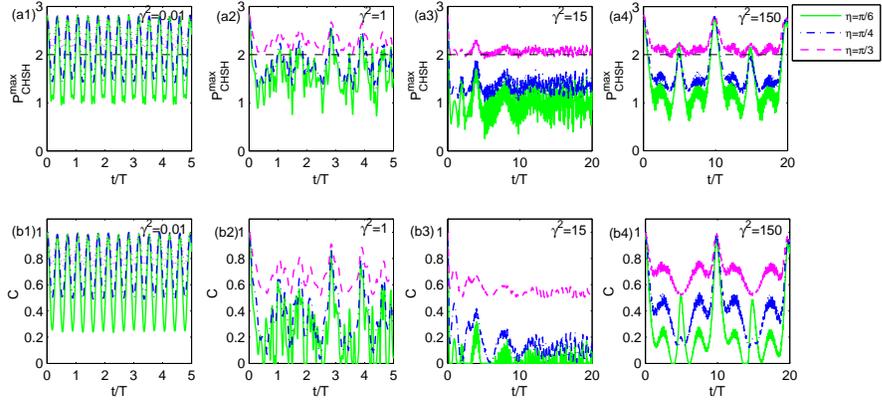}
\caption{{}Color online) Time-variation curves of $P_{CHSH}^{\max }(t)$ (a)
and $C(t)$ (b) for the entangled-state of antiparallel spin-polarization
with angle parameters $\protect\xi =\protect\pi /4$ and $\protect\eta =%
\protect\pi /6$ (green solid line), $\protect\pi /4$ (blue dot and dash
line) $\protect\pi /3$ (pink dash line) with average photon-number $\protect%
\gamma ^{2}=0.01$ (1), $1$ (2), $15$ (3), $150$ (4). }
\end{figure}

The time-variation curves of maximum CHSH-correlation-probability $%
P_{CHSH}^{\max }(t)$ (a) and the concurrence $C(t)$ (b) are displayed in
Figs. 1-2, for initial entangled states of antiparallel spin-polarizations
Eq. (\ref{p01}) with various average photon-numbers $\gamma ^{2}=0.01$ (1), $%
1$ (2), $15$ (3), $150$ (4). The time $t$ is measured in the optic-field
period $T=2\pi /\omega $ and the atom-filed coupling $g=1$ is in the unit of
field frequency $\omega $. The time-variation curves of $P_{CHSH}^{\max }(t)$
and $C(t)$ have one to one correspondence with the upper bound value $%
P_{CHSH}^{\max }=2\sqrt{2}$, which corresponds to the concurrence $C=1$
indicating the maximum entanglement. The dynamics is sensitive to the angle
parameters $\xi $ and $\eta $ of the entangled states. For $\eta =0$ the two
coefficients of the entangled states are real and the time-evolution curves
are plotted for the parameters $\xi =\pi /6$ (green solid line), $\pi /4$
(blue dot and dash line), and $3\pi /4$ (red dash line) in Fig. 1. It is
remarkable to see a fact that both the maximum CHSH-correlation-probability
and concurrence remain in the maximum values $P_{CHSH}^{\max }(t)=2\sqrt{2}$%
, $C(t)=1$ not varying with time when $\xi =3\pi /4$. In this case the
initial state becomes an eigenstate of the system $\left\vert \psi
_{1}\right\rangle $ i.e. the two-spin singlet state as shown in Eq. (\ref%
{Eq4}), which is independent of the cavity-field parameters $\gamma $ and $%
\phi $. The measuring outcome of two-spin correlation operator is invariant
in the Hamiltonian eigenstate $\left\vert \psi _{1}\right\rangle $, since
the density operator commutes with the Hamiltonian and thus is a conserved
quantity. If the initial state is not the eigenstate with $\xi =\pi /6$, $%
\pi /4$, $P_{CHSH}^{\max }(t)$ oscillates with time below the quantum
upper-bound $2\sqrt{2}$. The oscillation behavior depends on the average
photon numbers. In the lower value of average photon-number $\gamma
^{2}=0.01 $, $P_{CHSH}^{\max }(t)$ is a periodic oscillation around the
classical upper-bound value of $P_{CHSH}^{lc}=2$ seen in Eq. (\ref{lc}). The
oscillation becomes random for $\gamma ^{2}$ tending to $1$. When the
average photon-number increases to $\gamma ^{2}=15$, the entanglement
collapses seen from Fig. 1(b3), since we have $P_{CHSH}^{\max }(t)$ $<2$,
namely the CHSH inequality Eq. (\ref{lc}) is satisfied. In other words the
quantum interference part $\rho _{\psi }^{nlc}$ disappears due to the
field-interaction induced decoherence. In the large photon-number limit the
coherence revives periodically characterized by the sharp-peak values, which
approaches the quantum upper-bound $2\sqrt{2}$ $(1)$ seen from Fig.
1(a4,b4). The discrete peak-positions of $P_{CHSH}^{\max }(t)$ curves are
located precisely at $t=n5T$ with $n$ being a integer. When the
superposition coefficients of entangled states become complex with
non-vanishing $\eta $, the periodic coherence-revival basically remains as
shown in Fig. 2 for $\xi =\pi /4$ and various phase angles $\eta $ for small
or large photon-number limit. It is interesting to see a fact that the CHSH
inequality is nearly always violated independent of average photon numbers
for the state with angle parameter $\eta =\pi /3$ (pink dash line), although
the $P_{CHSH}^{\max }(t)$ value is suppressed from the quantum upper-bound.
The field interaction does not lead to the decoherence for this state. While
the entanglement collapses for the states with $\eta =\pi /4$, $\pi /6$ at
the average photon number $\gamma ^{2}=15$ (b3). The sharp peak-positions of 
$P_{CHSH}^{\max }(t)$ curves become $t=(2n)5T$ in the large photon-number
limit with $\gamma ^{2}=150$ (a4). While The peak-height almost shrinks to
the classical bound $2$ at the time points $t=$ $(2n+1)5T$. The concurrence
plots are displayed in (b1) to (b4) showing the same behaviors as $%
P_{CHSH}^{\max }(t)$.

\subsection{Entangled states for parallel spin polarizations}

The BI was originally formulated and well studied based on the two-spin
singlet state $\left\vert \psi _{1}\right\rangle $. It was proved long ago 
\cite{pla} and verified recently by means of quantum statistics \cite{Song}
that the BI and its violation are actually valid for arbitrary two-spin
entangled states with antiparallel polarization \cite{pla,Song} given by Eq.
(\ref{p01}). Most recently it was demonstrated that the original formula of
BI has to be slightly modified for the two-spin entangled state of parallel
spin polarization, while the CHSH inequality remains not changed \cite%
{Zhang4}. Now we turn our attention to the dynamics of the maximum
quantum-CHSH-probability $P_{CHSH}^{\max }(t)$ and the concurrence $C(t)$
for the initial entangled states with parallel spin-polarizations given by
Eq. (\ref{bbbb}). The $P_{CHSH}^{\max }(t)$ and $C(t)$ curves are displayed
respectively in upper (a) and lower (b) panels in Fig. 3 for the average
photon-numbers $\gamma ^{2}=0.01$ (1), $5$ (2), $15$ (3), $150$ (4) with the
state parameters $\eta =0$, $\xi =\pi /3$ (green solid-line), $\pi /4$ (blue
dot and dash line), $3\pi /4$ (red dash-line). Again the time evolution
becomes random when the average photon number reaches $\gamma ^{2}=1$. When
the average photon-number increases to $\gamma ^{2}=15$, the entanglement
collapses only for the initial state with $\xi =3\pi /4$. The CHSH
inequality is, however, violated for the states with $\xi =\pi /6$, $\pi /4$
since $P_{CHSH}^{\max }(t)>2$ in these cases. $P_{CHSH}^{\max }(t)$ remains
in the upper quantum bound value with small oscillation for the state $\xi
=\pi /4$ (blue dot-and dash-line), while the average value of $%
P_{CHSH}^{\max }(t)$ for the state $\xi =\pi /6$ (green solid line) is
slightly lower than the upper bound. In the large photon-number limit $%
\gamma ^{2}=150$ (a4), the decoherence for the initial state $\xi =3\pi /4$
revivals periodically at the time points $t=n5T$. Correspondingly the
time-variation curves of $P_{CHSH}^{\max }(t)$ and $C(t)$ are displayed in
Fig. 4 for the initial states with complex superposition coefficients, in
which the angle parameters are chosen as $\xi =\pi /4,$ $\eta =\pi /6$
(green solid-line), $\pi /4$ (blue dot- and dash-line), $\pi /3$ (pink
dash-line) respectively. At the average photon-number $\gamma ^{2}=15$, the
entanglement collapses for the initial states with $\eta =\pi /4$, $\pi /3$
(b3) and revivals at the time points of $t=(n/2)5T$ in the large
photon-number limit $\gamma ^{2}=150$ (b4). However, the peak-height of $%
P_{CHSH}^{\max }(t)$ can approach the upper quantum bound only at $t=(2n)5T$%
. The $P_{CHSH}^{\max }(t)$-curve varies with time for the state $\eta =\pi
/6$ above the classical bound and so that the coherence remains.

\begin{figure}[ptb]
\centering\includegraphics[width=13cm]{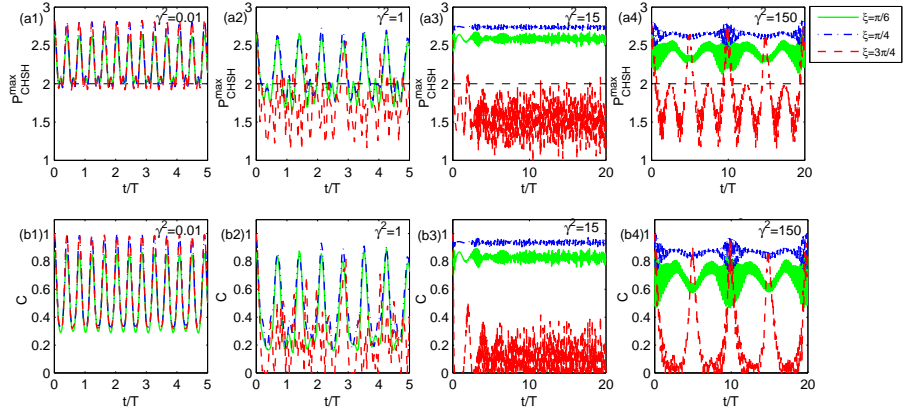}
\caption{{}Color online) The maximum CHSH correlation-probability $%
P_{CHSH}^{\max }(t)$ (upper panel) concurrence $C(t)$ (lower panel) as
functions of time $t$ in the initial entangled states of parallel
spin-polarizations with angle parameters $\protect\eta =0$, $\protect\xi =%
\protect\pi /3$, $\protect\pi /4$, and $3\protect\pi /4$ for the average
photon-number $\protect\gamma ^{2}=0.01$ (1), $1$ (2), $15$ (3), $150$ (4). }
\end{figure}

\begin{figure}[ptb]
\centering\includegraphics[width=13cm]{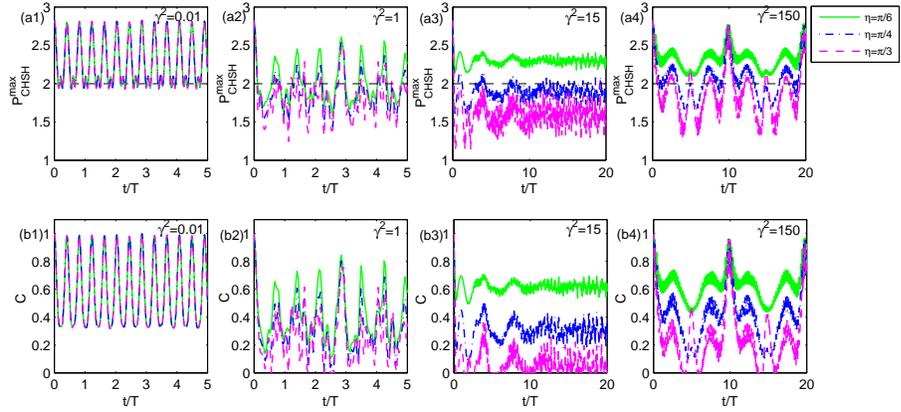}
\caption{{}Color online) The time variation of $P_{CHSH}^{\max }(t)$ (upper
panel) and concurrence $C(t)$ (lower panel) in the entangled states of
parallel spin-polarizations with the angle parameters $\protect\xi =\protect%
\pi /4,$ $\protect\eta =\protect\pi /6$, $\protect\pi /4$, and $\protect\pi %
/3$ for the average photon-number $\protect\gamma ^{2}=0.01$ (1), $1$ (2), $%
15$ (3), $150$ (4). }
\end{figure}

.

\section{Conclusion and discussion}

We propose a theoretical model for two spins in a single-mode optical cavity
with the ordinary magnetic interaction. Stationary entangled-states are
realized in terms of variational method under the condition of optical
coherent-state. The well studied two-spin singlet state is indeed generated
in this simple model. Besides we also obtain the entangled state with
parallel spin polarizations. The BI and its violation are reformulated in a
unified way by means of quantum probability statistics with the help of
state density operator. This formulation establishes a direct relation
between decoherence and BI, which is obtained by neglecting the quantum
interference part of density operator. Thus the BI indeed can serve as a
criteria of decoherence. The BI in the CHSH form is also valid for the
general two-spin entangled states of parallel spin polarizations so that the
entanglement dynamics can be investigated for arbitrary entangled states
with both antiparallel and parallel spin polarizations. Based on the
time-evolution of state-density operator derived with the quantum master
equation the dynamics of maximum CHSH-correlation-probability $%
P_{CHSH}^{\max }(t)$ is evaluated in comparison with the concurrence $C(t)$
of two-spin entangled states. It is interesting to see a fact that the
maximum quantum CHSH-probability as well as the concurrence does not vary
with time in the spin singlet state. This is because that the density
operator, which commutes with Hamiltonian, is a conserved quantity, and the
measuring outcome is invariant with time in this state. In the general
entangled states with antiparallel spin polarizations, the time-evolutions
of $P_{CHSH}^{\max }(t)$ and $C(t)$ are periodic in the weak field regimen
with the average photon number in the order of $\gamma ^{2}=0.01$. The spin
flip due to the field interaction leads to the time oscillation of measuring
outcome correlation. When the field intensity increases to the order of $%
\gamma ^{2}=1$ the time evolution becomes random. The entanglement collapses
at the average photon number $\gamma ^{2}=15$ due to the field induced
decoherence, which is characterized by the fulfilling of CHSH-inequality.
The coherence revivals periodically in the large photon-number limit $\gamma
^{2}=150$, such that the sharp peaks of $P_{CHSH}^{\max }(t)$ curves
approach the upper quantum-bound $2\sqrt{2}$. The peak positions are located
precisely at $t=n5T$. For the complex superposition-coefficients of the
entangled states the complete decoherence appears for the states $\eta =\pi
/6$, $\pi /4$. Coherence revivals at the time points $t=(2n)5T$, but not at $%
t=(2n+1)5T$ , where the peak heights shrink to the classical bound value $2$%
. The entanglement dynamics is similar but different in details for the
states with parallel spin-polarizations. The precise period of coherence
revival associated with the cavity-field frequency is an interesting
observation in the large photon-number limit.

\section*{Acknowledgments}

This work was supported by the Natural Science Foundation of China (Grants
Nos. 11105087, 61275210, 11275118, 11404198) and Scientific and
Technological Innovation Programs of Higher Education Institutions in Shanxi
(STIP) (Grant No. 2014102).

\end{document}